\documentclass[conference,12pt,onecolumn]{IEEEtran}

\usepackage{fullpage,graphicx,psfrag,epsfig,amsmath,amsfonts}
\usepackage[dvips]{color}

\newtheorem{theorem}{Theorem}

\newtheorem{lemma}{Lemma}

\newtheorem{conjecture}{Conjecture}

\newcommand{\enp} {\hfill \rule{2.2mm}{2.6mm}}

\begin{document}
\pagestyle{empty}
\title{Towards the distribution of the smallest matching in the Random
Assignment Problem}

\author{\authorblockN{Chandra Nair}
\authorblockA{Dept. of EE\\
Stanford University\\
Stanford, CA 94305\\
Email: mchandra@stanford.edu}}

\maketitle

\begin{abstract}
We consider the problem of minimizing cost among one-to-one
assignments of $n$ jobs onto $n$ machines. The random assignment
problem refers to the case when the cost associated with
performing jobs on machines are random variables. Aldous
established the expected value of the smallest cost, $A_n$, in the
limiting $n$ regime. However the distribution of the minimum cost
has not been established yet. In this paper we conjecture some
distributional properties of matchings in matrices. If this
conjecture is proved, this will establish that $\sqrt{n}(A_n -
E(A_n)) \overset{w}{\Rightarrow} N(0,2)$. We also establish the
limiting distribution for a special case of the Random Assignment
Problem.
\end{abstract}

\IEEEpeerreviewmaketitle

\section{Introduction}
Consider the problem of assigning $n$ jobs onto $n$ machines. Let
$c_{i,j}$ denote the cost of performing job $i$ on machine
$j$. Consider a 1-1 assignment, $\pi$, that matches the $n$ jobs to
$n$ machines. Let
\[ A_n^\pi = \sum_{i=1}^n c_{i,\pi(i)} \]
denote the cost associated with the matching (assignment)
$\pi$. Further let $\pi^*$ denote the matching that minimizes the cost
among all $n!$ matchings. Define
\[A_n = \sum_{i=1}^n c_{i,\pi^*(i)} \]
It has been shown that when the random variables $c_{i,j}$ are
independent and identically distributed, the distribution of
$A_n$ is only dependent on the value of the density function at
origin. Two popular choices in literature for the density function of
$c_{i,j}$ have been $U[0,1]$ and $\exp(1)$. We will assume throughout
the rest of the paper that $c_{i,j}$ are i.i.d. exp(1) random
variables.

\subsection{Recall of some results}

For large values of $n$, Lazarus \cite{laz} showed that $E(A_n) >
1+\frac{1}{e}$. Later, Olin \cite{olin} improved it to 1.51 in her
Ph.D. thesis. Walkup, \cite{walkup}, established an upper bound of 3
for $E(A_n)$ in the large $n$ regime. In \cite{karp84} Karp improved
this upper bound from 3 to 2. Coppersmith and Sorkin improved
this bound further to 1.91 in \cite{cas}.

Using Replica Method, a technique developed by Statistical
Physicists to study interactions between particles, Mezard and
Parisi, argued that the limit of $E(A_n)$ was $\frac{\pi^2}{6}$.
They also computed the distribution of a randomly chosen entry
that was part of the smallest matching. However this method makes
assumptions that cannot be rigorously justified. They also claimed
that the assignment problem had the 'self averaging property',
i.e. the distribution of $A_n$ concentrates around the mean for
large $n$.

In \cite{aldous92}, Aldous rigorously established that the limit
of $E(A_n)$ exists. Later, in \cite{aldous01}, he established that
the limit was $\frac{\pi^2}{6}$, as predicted by the physicists.
He also recovered the distribution for a random entry in the
smallest assignment. As a further evidence to the Physicists'
approach, Talagrand showed that the variance decayed at a rate
that was lower bounded by $\frac{1}{n}$ and upper bounded by
$\frac{\log^4{n}}{n}$.

\subsection{Finite Random Assignment Problem}

For every finite $n$, Parisi conjectured that
\[ E(A_n) = \sum_{i=1}^n \frac{1}{n^2} \]
This was established last year simultaneously using very different
approaches in \cite{lw03} and \cite{nps03}. The latter approach
is an size-based induction on matchings building from the smallest
entry (smallest matching of size 1) to the smallest matching of size
$n$.

In \cite{nps03}, the authors establish the distribution of these
increments from the smallest matching of size $k$ to the smallest
matching of size $k+1$. Though the smallest matching of size $n$
is the sum of these increments, correlations between these random
variables prevent them from getting the distribution for the
smallest matching of size $n$. However linearity of the
expectation was sufficient for them to get the expected value of
the smallest matching of size $n$. In this paper, we conjecture
the exact nature of these correlations in the large $n$ regime.
These conjectures imply that
\[ \sqrt{n}(A_n - E(A_n)) \overset{w}{\Rightarrow} N(0,2)\]

\subsection{Results on the limiting distribution}

In \cite{aldous01}, Aldous commented that one would expect the
limiting distribution to be Gaussian. In \cite{almso02}, Alm and
Sorkin conjectured that the limiting variance of $\sqrt{n}(A_n -
E(A_n))$ is 2. The basis of the conjecture regarding the variance,
according to the authors, is based on a communication between Janson
and the authors in which Janson guessed the exact distribution for
every finite $n$. This guess turned out to be incorrect for $n \geq 3$
but seemed very close to the true distribution.

The conjecture in this paper regarding the correlations in the
large $n$ regime, when applied to finite $n$ will yield
distributions that have a lot of similar terms to that of Janson's
guess. However, the finer nature of our conjectures and the
differences in some terms help us conclude that the limiting
distribution is Gaussian rather easily.

In the last section of this paper, we consider a special cost
matrix in which $n-1$ diagonal entries are zero and the rest are
i.i.d. exp(1) entries. We identify the scaled limiting
distribution for this case. The limiting distribution follows from
a very simple case of a theorem in \cite{na04} and a connection
between the form of the distribution to the distribution of
shortest path lengths in complete graphs.

\subsection{Notation and Recall}

Consider an $n \times n$ matrix, $M$. For any $k$ such that $1
\leq k \leq n-1$, let $T_1^k$ represent the smallest matching of
size $k$ in this matrix. A matching is a collection of elements
with the property that no two elements lie in the same row or same
column. Note that $T_1^k$ occupies some $k$ columns of $m$ and
w.l.o.g we can assume it is the first $k$ columns. Now let $S_i^k$
denote the smallest matching of size $k$ in the $n \times n-1$
matrix obtained from $M$ by the removal of column $i$. Thus one
obtains the $k+1$ matchings $T_1^k, S_1^k, ... , S_k^k$. (Observe
that removal of any column outside the first $k$ will yield
$T_1^k$ as the smallest matching).

Sort the matchings in order of their increasing weight to obtain
the sequence $T_1^k,...,T_{k+1}^k$. (Note: $T_1^k$ being the
smallest matching of size $k$ in the entire matrix will be smaller
than every $S_i^k$). In a slight deviation from the notation in
\cite{nps03}, let $t_i^k$ denote the weight of the matching
$T_i^k$. Define $T_1^n$ to be smallest matching $\pi^*$ and hence
$t_1^n = A_n$.

We recall the following result from \cite{nps03}. \vspace{0.05in}
\begin{theorem}
\label{thm1}
The following hold:
\begin{itemize}
\item $t_{i+1}^k - t_i^k \sim \exp(n-i+1)(n-k+i-1)$
\item $\{t_2^k - t_1^k, t_3^k - t_2^k,..., t_{k+1}^k-t_k^k,
 t_1^{k+1}-t_{k+1}^k\}$ are independent
\end{itemize}
\end{theorem}

\vspace{0.1in}
Theorem 1 gives an explicit characterization of the distribution
relating the difference between the smallest matching of size $k+1$
and the smallest matching of size $k$ in terms of sums of independent
exponentials.

{\em Remark:} Note that Theorem 1 does not give the entire
distribution as it does not say anything regarding the dependence of
the variables $t_1^{k+1} - t_1^k$ and $t_1^k - t_1^{k-1}$.

\section{Conjectures on Correlation}
Consider a set of variables $\Delta_i^k$ defined by the following set of
equations. All the random variables are assumed to be
independent.
\begin{eqnarray*}
\label{struc}
\Delta_1^k &\sim& \exp(n(n-k+1)), ~\mbox{for}~ k=1,...,n\\
\Delta_i^k & \sim & \left\{ \begin{array}{lll} 0 & \mbox{w.p.} & \frac{n-i+1}{n-i+2} \\
\exp(n-i+1)(n-k+i) & \mbox{w.p.} & \frac{1}{n-i+2}
\end{array} \right. ~ {\scriptstyle 2 \leq i \leq k ~,~ 2 \leq k \leq
n } \\
\end{eqnarray*}

Now define random variables $r_i^k$ recursively according to the
following relations:
\begin{eqnarray*}
r_1^1 & = & \Delta_1^1 \\ r_2^k - r_1^k & = & \Delta_1^{k+1}
~~~~~~~~~~~~~~~~~~~\mbox{for}~ k=1,...,n-1 \\ r_1^2 - r_2^1 &=& r_1^1
+ \Delta_2^2 \\ r_{i+1}^k - r_i^k &=& r_i^{k-1} - r_{i-1}^{k-1} +
\Delta_i^{k+1} \:~\mbox{for}~ i=2,..,k \\ r_1^{k+1}-r_{k+1}^k &= & r_1^k
- r_k^{k-1} + \Delta_{k+1}^{k+1} ~~~~\mbox{for}~ k=2,..,n-1
\end{eqnarray*}

It is easy to see that $r_i^k$'s satisfy the conditions of Theorem
\ref{thm1}, i.e.
\begin{itemize}
\item $r_{i+1}^k - r_i^k \sim \exp(n-i+1)(n-k+i-1)$
\item $\{r_2^k - r_1^k, r_3^k - r_2^k,..., r_{k+1}^k-r_k^k,
 r_1^{k+1}-r_{k+1}^k\}$ are independent
\end{itemize}

Observe that this equivalence of the marginals of the increments also
implies $E(t_i^k) = E(r_i^k)$.

{\em Remark:} The initial guess was that the distribution of
$r_i^k$ was in fact the distribution of $t_1^k$. However this was
observed not to be true for $n\geq3$. Calculations for $n=3$ and
$n=4$ demonstrated that the distribution of $r_i^k$ and $t_i^k$
are very close to each other though not exactly equal. Simulations
for higher $n$ confirm this observation. This makes us conjecture
that the under the correct scaling (i.e. multiplication by
$\sqrt{n}$) the error terms are of lower order and they die down
as $n$ becomes large.

\vspace{0.1in}
\begin{conjecture}
\label{mainconj} Let $F_n(x) = \mathbb{P}[\sqrt{n}(t_i^k -
E(t_i^k)) \leq x]$ and let $G_n(x) = \mathbb{P}[\sqrt{n}(r_i^k -
E(tr_i^k)) \leq x]$. Then $|F_n(x)-G_n(x)| \rightarrow 0, \forall
x$  as $n \rightarrow \infty$.
\end{conjecture}
\vspace{0.1in}

Assuming that Conjecture 1 is correct, then this would imply that if
\begin{eqnarray*} \sqrt{n}(r_1^n - E(r_1^n))
&\stackrel{w}{\Rightarrow}& N(0,2), ~\mbox{then} \\
\sqrt{n}(A_n - E(A_n)) &\stackrel{w}{\Rightarrow}&  N(0,2),
~\mbox{since}~  t_1^n = A_n
\end{eqnarray*}
We prove the first claim in the lemma below.

\vspace{0.1in}
\begin{lemma}
\label{conv}
$\sqrt{n}(r_1^n - E(r_1^n)) \stackrel{w}{\Rightarrow} N(0,2)$.
\end{lemma}

\noindent\proof Writing $r_1^n$ in terms of the random variables $\Delta_i^k$
we obtain the following relation.
\[ r_1^n = \sum_{k=1}^n \sum_{i=1}^{k} (n-k+1)\Delta_i^k \]
Let $\mu_i^k$ = E($\Delta_i^k$) and let $\mu_n = E(r_1^n)$. Then we
note the following:
\begin{eqnarray}
&&\lim_{n} n(E(r_1^n - \mu_n)^2) = \lim_n n\sum_{k=1}^n \sum_{i=1}^{k}
(n-k+1)^2 E(\Delta_i^k - \mu_i^k)^2  = 2 \label{var}\\
&&\lim_{n} n^2  \sum_{k=1}^n\sum_{i=1}^{k} (n-k+1)^4 E(\Delta_i^k - \mu_i^k)^4
= 0 \label{four}
\end{eqnarray}

The proofs of these two equations were obtained using {\em
MATHEMATICA} and hence has been omitted from the paper.

Now we apply the Central Limit Theorem for arrays to finish the
argument.  Let $X_{n,k,i} = \sqrt{n}(n-k+1)(\Delta_i^k -
\mu_i^k)$.  Observe that $\sum_{k,i} X_{n,k,i} = \sqrt{n}(r_1^n -
E(r_1^n))$.

Eqns (\ref{var}) and (\ref{four}) imply the following
conditions for the zero-mean independent random variables $X_{n,k,i}$.
\begin{itemize}
\item $\lim_{n} \sum_{k,i}E(X_{n,k,i}^2) = 2.$
\item $\lim_n \sum_{k,i}E(X_{n,k,i}^4) = 0.$
\end{itemize}
Hence they satisfy the Lyapunov conditions for CLT and thus we have
\[ \sum_{k.i} X_{n,k,i}  \stackrel{w}{\Rightarrow} N(0,2) ~\mbox{as}~ n
\rightarrow \infty \]
This completes the proof of the lemma and hence assuming Conjecture 1
is true, this establishes the limiting distribution of $A_n$. \enp

{\em Remark:} Though the Lyapunov CLT is normally stated with the
third moment rather than the fourth moment used here, it is easy
to see that any 2 + $\delta$ moment is sufficient.)

Now consider the increment $r_1^{k+1} - r_1^k$. The distribution for
this increment can be explicitly stated in terms of sums of
independent exponentials as stated in Theorem 1. However, from the
definition of the random variables $r_i^k$ we get the following
relation:
\[ r_1^{k+1} - r_1^k = r_1^k - r_1^{k-1} + \sum_{i=1}^k \Delta_i^{k+1} \]

Hence $r_1^{k+1}- r_1^k > r_1^k - r_1^{k-1}$. The following lemma
shows that this is true for $t_i^k$'s also.

\begin{lemma}
\label{diff} $t_1^{k+1} - t_1^k > t_1^k - t_1^{k-1}$
\end{lemma}
\proof Re-arranging the terms it is sufficient to show that $t_1^{k+1}
+ t_1^{k-1} > 2t_1^k$.

{\em Case 1:} If the matching $T_1^{k+1}$ contains one element
that lies outside the rows and columns occupied by $T_1^{k-1}$,
then we can combine this element with the matching $T_1^{k-1}$ and
get a matching of size $k$. Note that the rest of the elements of
$T_1^{k+1}$ is a matching of size $k$. Therefore we can identify
two matchings of size $k$ from among the elements of $T_1^{k-1}$
and $T_1^{k+1}$. Therefore the combined weight of these two
matchings of size $k$ must be greater than twice the weight of the
smallest matching of size $k$.

{\em Case 2:} When there is no element of $T_1^{k+1}$ that lies
outside the rows and columns of $T_1^{k-1}$ we establish the lemma
by using the following two properties of matchings. First, the
rows and columns used by the smallest matching of size $k$
contains all the rows and columns used by the smallest matching of
size $k-1$.  We represent a matching as a bipartite graph and an
edge is present between node $i$ on the left and node $j$ on right
if the element $(i,j)$ is present in the matching. The second
property is that we can decompose two matchings (represented on
the same bipartite graph) into the following three components:
common edges, alternating paths and alternating cycles.

Consider a bipartite graph formed by the elements of $T_1^{k+1}$
and $T_1^k$. From the first property this is a $k+1 \times k+1$
bipartite graph. Color the $k+1$ edges represented by the elements
of $T_1^{k+1}$ by red and the $k-1$ edges represented by the
elements of $T_1^{k-1}$ by green. Now from the minimality of these
matchings there cannot be any cycles. The first property also
implies that the alternating paths must be of odd length and must
have one extra red edge. (If it is of even length or has one extra
green edge then we see that property one is violated). Therefore,
we can decompose the bipartite graph into common edges and two
alternating paths each having one extra red edge.

Now form one matching of size $k$ by picking the common edges, red
edges from first alternating path and green edges from second
alternating path. Form the second matching of size $k$ by picking
common edges, green edges from first alternating path and red edges
from second alternating path. Observe that the total weight of these
two matchings of size $k$ is equal to $t_1^{k+1} + t_1^{k-1}$. But
this should be greater than twice the weight of the smallest matching
of size $k$. This completes the proof of the lemma for Case 2. \enp

\section{Limiting Distribution for a special case}

In this section, we assume that the cost matrix has the following
form. $C_{i,i} = 0$ for $1 \leq i \leq n-1$. The rest of the
entries are assumed to be i.i.d. $\exp(1)$ random variables. Let
$\tilde{A}_n$ represent the weight of the minimum matching when
the cost matrix has this structure.

We recall some notation from \cite{nps03}. Consider a $(n-1)
\times n$ matrix of i.i.d. exp(1) entries. Let $T_1$ be the
smallest matching of size $n-1$. W.l.o.g. let it occupy the first
$n-1$ columns. Let $S_i$ be the smallest matching of size $n-1$ in
the $(n-1) \times (n-1)$ matrix obtained after deleting the $i$th
column. This gives the matchings $T_1, S_1,...,S_{n-1}$. Now
arrange them in increasing order of their weights to obtain the
sequence $T_1,T_2,...,T_n$. Further, as before let $t_i$ denote
the weight of the matching $T_i$.

We recall the following theorem from \cite{nps04}.
\begin{theorem}
\label{row} Consider a $n-1 \times n$ matrix of i.i.d. exp(1)
entries. Now conditioned on a fixed placement of minimum in each
row the following hold:
\begin{itemize}
\item $t_{i+1}-t_i \sim \exp(i(n-i))$
\item $\{t_2-t_1,...,t_n - t_{n-1} \}$ are
independent.
\end{itemize}
\end{theorem}

{\em Remark:}The proof of this theorem follows the same line of
argument as the proof for the case without conditioning on the
placement of the minima that is in \cite{nps03}. However the
details of the argument is in \cite{na04}. Note that the special
case of the theorem \ref{row} used below is very straightforward
and can be established without the machinery of \cite{nps03}.

Consider a special case of Theorem \ref{row} where we assume that
all the minimum lie in different rows. Now we form a new matrix by
subtracting the minimum entry in each row from all the entries in
the row. By the memoryless property of the exponential
distribution, this new matrix reduces to having zeroes where the
minimum in each row was present and i.i.d. exp(1) random variables
in other locations. W.l.o.g. we can assume that the zero entries
are located at $(i,i)$ for $i = 1,..,n-1$. Also observe that for
this new matrix, the weights of its $T-$matchings are
$\{0,t_2-t_1,...,t_n-t_1\}$ where $t_i$ are the weights of the
$T-$matchings of the original matrix.

Now observe that $\tilde{A}_{n-1}$ has the same distribution as
$s_{n-1}-t_1$. However by symmetry, this is equally likely to be
one of ($t_2 - t_1,t_3-t_1,...,t_n-t_{n-1}$). Therefore from
Theorem \ref{row} it follows that the distribution of
$\tilde{A}_{n-1}$ is given by:
\begin{equation}
\label{sur}
\tilde{A}_{n-1} \sim \left\{
\begin{array}{rcl} \exp(n-1) & w.p. & \frac{1}{n-1} \\ \exp(n-1) +
\exp2(n-2) &w.p.& \frac{1}{n-1}
\\ & & \\ & & \\ \sum_{k=1}^{n-1} \exp(k(n-k)) &w.p.& \frac{1}{n-1} \end{array} \right.
\end{equation}

Consider the following unrelated problem. There is a complete
graph on $n$ vertices and its edge weights are i.i.d. exp(1)
random variables. Let $i$ and $j$ be two randomly chosen vertices.
Let $X_{ij}^n$ denote the weight of the cheapest path from $i$ to
$j$. Then the distribution of $X_{ij}^n$ is given by eqn.
(\ref{sur}). This was shown by Janson in \cite{jan99}. However a
deeper connection between the problems other than the algebraic
equivalence of the distributions has proved elusive.

In \cite{jan99}, Janson also computes the asymptotic distribution
of the random variable $X_{ij}^n$. He shows that \[ X_{ij}^n -\log
n \stackrel{w}{\Rightarrow} W_1 + W_2 - W_3
\] where $W_1, W_2, W_3$ are independent random variables with the
same extreme value distribution $P(W_i \leq x) = e^{-e^{-x}}$.

In \cite{al04p}, Aldous had suggested that the limiting
distribution of the random variable governing the asymptotic
distribution of $X_{ij}^n$ could possibly be written as
\[ nX_{ij}^n -\log n \stackrel{w}{\Rightarrow} C \] where
$P(C > x)= E(e^{-UVe^x})$, where U,V are independent exp(1) random
variables. The equivalence of Janson's result and the form
suggested by Aldous is quite straightforward.

Thus, the limiting distribution for the special cost matrix in
which all entries on the diagonal except one are zero and the rest
of the entries are i.i.d. exp(1) entries is given by
\[n\tilde{A}_n - \log n \stackrel{w}{\Rightarrow} C \] where the
random variable $C$ is as defined above.

\section{Conclusion}
In this paper we conjecture that the increments of the small
matchings in a matrix has a particular correlation structure in
the large $n$ regime. This conjecture, if true, would prove that
the limiting distribution of the smallest matching when properly
scaled and centered would converge to N(0,2).

A similar set of conjectures can also be stated for the case where
the matrices are rectangular and using those correlation structure
one could guess the limiting distribution for the case for an $m
\times n$ matrix. Note however one expects the right scaling to be
proportional to $\sqrt{n}$ only if $m$ scales as $ \alpha n$ for
some $\alpha > 0$.

We also prove the limiting distribution for a special case of the
cost matrix. However this is based on a algebraic identity between
the distributions in two seemly unrelated problems. It would be
good to understand if it is something more than a coincidence.

\section*{Acknowledgment}

The author would like to thank Mohsen Bayati and Balaji Prabhakar
for patiently listening to the various approaches attempted by the
author and giving useful feedback. The author also thanks  Mohsen
Bayati for bringing to attention Alm and Sorkin's conjecture on
the variance. The author would also like to thank Prof. Aldous and
Dr. Lovasz for suggestions and references that were very helpful.
Finally, the author would like to thank Sam Kavusi who helped in
the calculations using {\em MATHEMATICA}.

\nocite{}
\bibliographystyle{IEEEtran}
\bibliography{biblio}

\end{document}